\documentclass[12pt,aps,prd,preprint,tightenlines,
   showpacs,nofootinbib]{revtex4}
\newcommand{\PRE}[1]{{#1}} 

\usepackage{bm} \usepackage{epsfig}
\usepackage{slashed}
\usepackage{multirow}

\newcommand{\eqref}[1]{Eq.~(\ref{#1})}

\newcommand{\bea}{\begin{eqnarray}}
\newcommand{\eea}{\end{eqnarray}}

\begin{document}

\preprint{UCI-TR-2010-04}

\title{
\PRE{\vspace*{1.5in}} Instability of (1+1) de Sitter space in the presence of interacting fields
\PRE{\vspace*{0.3in}} }

\author{Myron Bander}
\affiliation{Department of Physics and Astronomy, University of
California, Irvine, CA 92697, USA \PRE{\vspace*{.5in}} }

\date{March  2010}

\begin{abstract}
\PRE{\vspace*{.3in}} 
Instabilities of two dimensional (1+1) de Sitter space induced by interacting fields are studied. As for the case of flat Minkowski space, several interacting fermion models can be translated into free boson ones and vice versa.  It is found that interacting fermion theories do not lead to any instabilities, while the interacting bosonic sine-Gordon model does lead to a breakdown of de Sitter symmetry and to the vanishing of the vacuum expectation value of the S matrix. 
\end{abstract}

\pacs{04.62.+v, 98.80.-k,11.10.Kk} 

\maketitle

\section{Introduction}
According to our present understanding and parametrization of cosmology, the universe is approaching a state described by a de Sitter metric.  Formulating quantum field theory on such a metric is, however, problematic \cite{{Goheer:2002vf},{Einhorn:2002nu}}. Recently Polyakov \cite{{Polyakov:2009nq},{Polyakov:2007mm}}
made the suggestion that interacting fields on such a background induce an instability and the positive curvature will decay, i.e. these interactions cause massive particle production that ultimately neutralizes the cosmological constant responsible for this curvature.  In \cite{Polyakov:2009nq} the effects of an interacting , $\lambda\phi^4$ massive scalar field were considered to order $\lambda^2$. It was found that the de Sitter symmetry is broken and that $\ln\langle0|S|0\rangle$ develops a large (proportional to the volume of space-time) negative real part, signaling a vacuum instability.  

In this work we will study this problem for the case of a (1+1) dimensional de Sitter background.  In flat (1+1) Minkowski space there are several interacting theories that can be solved exactly. Among these are: (i) the Thirring model \cite{Thirring}, (ii) massless QED \cite{Schwinger}, and (iii) spin-0 with a sine-Gordon interaction $\sim \cos(2\sqrt{\pi}\phi)\, .$  The reason these interacting field theories can be solved is that there is an equivalence \cite{{Coleman},{Kog-Suss}, {Bander:1975pi}}  wherein spinor fields can be written in terms of spin-0 ones and for the cases cited above the interacting theory is expressible as a free theory with opposite statistics. 

Such a correspondence between bosonic and fermionic formulations can be extended to a background de Sitter space.   The two interacting fermion models, (i) and (ii) above, go over to free spin-0 theories preserving de Sitter symmetry. No instability of de Sitter space is indicated. The case of bosons interacting by a sine-Gordon term, (iii) above, corresponds to a free, massive spin-$\frac{1}{2}$ field theory albeit with a mass term that depends on the de Sitter time, thus explicitly breaking de Sitter symmetry. A further analysis of this model shows that $\ln\langle0|S|0\rangle$ has an infinite real part. indicating a vacuum instability. 

In Section \ref{corres} this correspondence between spin-$\frac{1}{2}$ fermi fields and spin-0 bose ones is developed for the case of a background de Sitter space and a ``dictionary" for translating certain composite operators from one language to the other is set up.  in Section \ref{interact} this is explicitly applied to the interacting models discussed earlier and the interesting case of bosons interacting via a sine-Gordon term is worked out in greater detail in Section \ref{sine-gordon-interaction}.  A summary and discussion of the main results is given in Section \ref{summ}.

\section{Boson-fermion Correspondence in (1+1) de Sitter Space}\label{corres}
The procedure for translating the expectation values of products of fermion fields in a massless free fermion field theory on a de Sitter space to those of products of bose fields in a massless free boson theory on the same space will follow the one presented in \cite{Bander:1975pi} for Minkowski space.  For this purpose it is useful to use planar (or flat slicing) coordinates \cite{Spradlin:2001pw} where the expression for the metric is
\begin{equation}\label{dsmetric1}
ds^2=dt^2-e^{2Ht}dx^2\, ,
\end{equation}
and then to transform the above to conformal time
\begin{equation}\label{dsmetric2}
ds^2= \frac{d\tau^2-dx^2}{(H\tau)^2}\, ;
\end{equation}
the relation between $t$ and the conformal time $\tau$ is $-H\tau=\exp (-Ht)$. The utility of the above metric for setting up a boson-fermion correspondence is that the spatial coordinate $x$ ranges over $-\infty\le x\le +\infty$;  the corresponding conformal time $\tau$ ranges over $-\infty\le\tau\le 0$.  Fields, propagators and Lagrangians using (\ref{dsmetric2}) are conformally related to the corresponding expressions in flat Minkowski space \cite{Birrell:1982ix}.For fields these conformal transformations are
\begin{eqnarray}\label{conf_trans}
 \phi_M &\leftrightarrow &\phi_{dS}\ \ \ \ \ \ \ \ \ \ \ \ \ \ \ {\rm spin\  0}\, ;\nonumber\\
\psi_M &\leftrightarrow&\psi_{dS}/(H\tau)\ \ \ \ \ \ \ {\rm spin\  1/2}\, ;\\
A_{\mu;M}& \leftrightarrow & (H\tau)^2A_{\mu;dS}\ \ \ \ \ {\rm spin\  1}\, .\nonumber
\end{eqnarray}

The metric tensors implied by (\ref{dsmetric2}) are: $g_{0,0}=-g_{1,1}=(H\tau)^{-2}\, ,g_{0,1}=0$ with $\sqrt{-g}=(H\tau)^{-2}$; the corresponding {\it zweibeins}, $e^\mu_a$, which we need for a discussion of the spinor dynamics  are; $e^0_0=H\tau\, ,e^1_1=H\tau\, , e^0_1=e^1_0=0$. The connection tensor $\Gamma_\mu=0$. 
The action for a free, neutral, massive scalar field, $\phi$, is
\begin{equation}\label{free_lagr_0}
S_0=\frac{1}{2}\int d\tau dx \sqrt{-g}\left(g^{\mu\nu}\partial_\mu\phi\partial_\nu\phi-m_b^2\phi^2\right)=
  \frac{1}{2} \int d\tau dx \left(\partial_0\phi\partial_0\phi-\partial_1\phi\partial_1\phi-m_b^2\frac{\phi^2}{(H\tau)^2}\right)\, ;
\end{equation}
the one for a free massive spinor  $\psi$ is
\begin{eqnarray}\label{free_lagr_1/2}
 S_{\frac{1}{2}}& =&\int d\tau dx \sqrt{-g}\Bigg[ \frac{i}{2}
\left({\bar\psi}e^{\mu,a}\gamma_a\partial_\mu\psi-e^{\mu,a}\partial_\mu{\bar\psi}\gamma_a\psi\right)-m_f{\bar\psi}\psi
\Bigg]\nonumber\\
&= &\int d\tau dx \Big[ \frac{i}{2H\tau}\left({\bar\psi}\gamma_0\partial_0\psi-\partial_0{\bar\psi}\gamma_0\psi
-{\bar\psi}\gamma_1\partial_1\psi+\partial_1{\bar\psi}\gamma_1\psi\right)-m_f\frac{{\bar\psi}\psi}{(H\tau)^2}\Big]\, ;
\end{eqnarray}
and the one for a massless vector field $A_\mu$, in the gauge $A_1=0$ 
\begin{equation}\label{free_lagr_1}
S_1=\int d\tau dx\sqrt{-g} \frac{-1}{4}F_{\mu\nu}F_{\lambda\sigma}g^{\mu\lambda}g^{\nu\sigma}=-\int
d\tau dx \frac{(H\tau)^2}{2}(\partial_1A_0)^2\, .
\end{equation}
The conformal transformation in (\ref{conf_trans}) can be read of from the Lagrangian correspondences above.

Integration by parts permits us to rewrite the the right hand side of (\ref{free_lagr_1/2}) 
\begin{equation}\label{simp_free_lagr_1/2} 
S_{\frac{1}{2}}=\int d\tau dx \Bigg[\frac{i}{H\tau}\Big({\bar\psi}\gamma_0\partial_0\psi-{\bar\psi}\gamma_1\partial_1\psi
+\frac{1}{2\tau}{\bar\psi}\gamma_0\psi\Big)-m_f\frac{{\bar\psi}\psi}{(H\tau)^2}\Bigg]\, .
\end{equation}
From the above we note that the momentum canonical to $\psi$ is
\begin{equation}\label{spinor_pi}
\pi_\psi=\frac{\delta S_{\frac{1}{2}}}{\delta \partial_0\psi}=\frac{i}{H\tau}\psi^\dag\, ,
\end{equation}
implying the equal-$\tau$ anticommutation relation
\begin{equation}\label{anticomm}
\left\{\psi_a(\tau,x),\psi_b^\dag(\tau,y)\right\}=H\tau\delta(x-y)\delta_{ab}\, .
\end{equation}
\subsection{Fermi-bose field correspondence}
The expression for fermi fields in terms of bose ones in Ref. \cite{Bander:1975pi},  eq.(3.9), valid for Minkowski space together with (\ref{anticomm}) tells us what modification we need to make in order to obtain a similar relation valid for de Sitter space.
\begin {eqnarray}\label{sp_bos_rel}
\psi_1(\tau,x)&=&\left(\frac{\Lambda H\tau}{2\pi\gamma}\right)^{1/2}\exp[-i\sqrt{\pi}\Phi_+(\tau,x)]\nonumber\\
          &{}&\\
\psi_2(\tau,x)&=&\left(\frac{\Lambda H\tau}{2\pi\gamma}\right)^{1/2}\exp[-i\sqrt{\pi}\Phi_-(\tau,x)]\nonumber\, .
\end{eqnarray}
In the above $\Lambda$ is an ultra violet cut-off, $\gamma=0.577\cdots$ is the Euler–-Mascheroni constant and $\Phi_{\pm}$
depends on a free massless bose field $\phi(\tau,y)$,
\begin{equation}
\Phi_{\pm}=\int_{-\infty}^x dye^{y/R}[\partial _\tau\phi(\tau,y)\pm\partial_y\phi(\tau,y)\, ;
\end{equation}
$R$ is a spatial cutoff and the limit $R\rightarrow\infty$ will be taken at the end of all calculations.  
It is the factors $(H\tau)^{1/2}$ in front of the identities of (\ref{sp_bos_rel}) that distinguish this fermion -boson correspondence from the one in flat Minkowski space. 
\subsection{Composite Operators}\label{comp_oper}
Using (\ref{sp_bos_rel}) we obtain directly the translation of fermion mass operators into the language of bose fields
\begin {eqnarray}\label{mass_oper_1}
:{\bar\psi}\psi:&=&\frac{H\tau\Lambda}{\pi\gamma}\cos\big[2\sqrt{\pi}\int_{-\infty}^x dy e^{y/R}\partial_y,\phi(\tau,y)\big]\, ,\nonumber\\
&{}&\\
:{\bar\psi}\gamma_5\psi:&=&i\frac{H\tau\Lambda}{\pi\gamma}\sin\big[2\sqrt{\pi}\int_{-\infty}^x dy e^{y/R}\partial_y\phi(\tau,y)\big]\, .\nonumber
\end{eqnarray}
Bearing in mind the caveats expressed in Ref. \cite{Bander:1975pi}, it is convenient for comparing boson and fermion Lagrangians or actions  to set $R=\infty$ and obtain 
\begin{eqnarray}\label{mass_oper_2}
:{\bar\psi}\psi:&=&\frac{H\tau\Lambda}{\pi\gamma}\cos 2\sqrt{\pi}\phi(\tau,x)\, ,\nonumber\\
  &{}&\\
:{\bar\psi}\gamma_5\psi:&=&i\frac{H\tau\Lambda}{\pi\gamma}\sin 2\sqrt{\pi}\phi(\tau,x)\, .\nonumber
\end{eqnarray}
Again, it is the extra factors involving the conformal time $\tau$ that differentiate this correspondence from the one in flat space 
and it is these terms that will be responsible for breaking de Sitter symmetry for interacting theories.

We now turn to current operators. First we note that the Noether current and axial current obtained from (\ref{simp_free_lagr_1/2}) are 
\begin{eqnarray}\label{noth-curr}
j_\mu &=&\frac{1}{H\tau}:{\bar\psi}\gamma_\mu\psi:\, \nonumber\\
&{}&\\
j^5_\mu &=&\frac{1}{H\tau}:{\bar\psi}\gamma_\mu\gamma_5\psi\:, .\nonumber
\end{eqnarray}
This time the extra factors involving $\tau$ cancel and the correspondence is as in flat space.
\begin{eqnarray}\label{curr_oper}
j_\mu(\tau,x)&=&\frac{\epsilon_{\mu\nu}}{\sqrt{\pi}}\partial^\nu\phi(\tau,x)\, ;\nonumber\\
&{}&\\
j_\mu^5(\tau,x)&=&\frac{1}{\sqrt{\pi}}\partial_\mu\phi(\tau,x)\, .\nonumber
\end{eqnarray}

\section{Interacting Theories -- Correspondence}\label{interact}
We shall look at a class of two dimensional theories that, in one language, bose or fermi, have non-trivial interactions, while in the other language are free field theories.  These are: (i) the Thirring Model, (ii) massless fermion QED and (iii) a sine-Gordon interaction. 
\subsection{Massless Thirring model $\leftrightarrow$ Free massive boson}\label{thirring}The action for a fermion with a current-current interaction, Thirring model, on a de Sitter space is
\begin{equation}\label{thirr_action_fermi}
S_{\rm Thirring}=\int d\tau dx \Big[ \frac{i}{2H\tau}\left({\bar\psi}\gamma_0\partial_0\psi-\partial_0{\bar\psi}\gamma_0\psi
-{\bar\psi}\gamma_1\partial_1\psi+\partial_1{\bar\psi}\gamma_1\psi\right)-\frac{g}{2}(j_0j_0-j_xj_x\Big]\, ,
\end{equation}
which, using (\ref{curr_oper}) is equivalent to a free mass-less bose action with the fermi field --bose field identification (\ref{sp_bos_rel}) rescaled to
\begin{equation}
\psi_{1,2}=\left(\frac{H\tau\Lambda}{\pi\gamma}\right)^{1/2}\exp\left\{-i\sqrt{\pi}\int_{-\infty}^x  dy e^{y/R}\left[
   \partial_0\phi/\beta \pm\beta\partial_y\phi\right]\right\}\, ,
\end{equation}
and $\beta=(1+g/\sqrt{\pi})$. De Sitter symmetry holds in both formulations. 
\subsection{Massless QED}\label{qed}
With the photon field in the $A_1=0$ gauge, the fermi action is 
\begin{eqnarray}\label{qed_action}
S_{\rm QED}&=&\int d\tau dx \left[ \frac{i}{2H\tau}\left({\bar\psi}\gamma_0\partial_0\psi-\partial_0{\bar\psi}\gamma_0\psi
-{\bar\psi}\gamma_1\partial_1\psi+\partial_1{\bar\psi}\gamma_1\psi\right)-ej_0A_0+\frac{(H\tau)^2}{2}\left(\partial_1A_0\right)^2\right]\, . \nonumber\\
&{}&
\end{eqnarray}
Solving the equation of motion for $A_0$ an using (\ref{curr_oper}) results in a scalar field action as in (\ref{free_lagr_0}) with $m2_b=e^2/\pi$.  Again, the de Sitter symmetry is valid in both formulations. 

\subsection{Sine-Gordon Interaction}\label{sine_gordon}
We consider a $\cos\beta\phi$ interaction with a special value for $\beta$, namely $\beta=2\sqrt{\pi}$. 
\begin{equation}\label{sine_gordon}
S_{\rm sine-Gordon}=
  \frac{1}{2} \int d\tau dx \left[\partial_0\phi\partial_0\phi-\partial_1\phi\partial_1\phi-\frac{g}{(H\tau)^2}\cos\left(2\sqrt{\pi}\phi\right)\right]\, .
\end{equation}
Eq. (\ref{mass_oper_2}) allows us to identify the above with $S_{\frac{1}{2}}$ of ( \ref{simp_free_lagr_1/2}) with $m_f=g\pi\gamma/(H\tau\Lambda)$\, .
This explicit $1/\tau$ behavior of the fermion mass breaks de Sitter symmetry.We shall look at this case in greater detail in the next section.
\section{sine-Gordon interaction}\label{sine-gordon-interaction}
As was noted in the previous section, the spin-0 sine-Gordon action translates to a free, massive spin-$\frac{1}{2}$ theory, albeit with a mass that depends on the cosmic time. This, by itself, indicates a breaking of de Sitter symmetry.  In this section we will investigate what effects this has on vacuum to vacuum transition amplitudes. 

In the fermionic language the action is 
\begin{equation}\label{taumass}
S_{\tau-{\rm dep-mass}}\int d\tau dx \frac{i}{H\tau}\Big({\bar\psi}\gamma_0\partial_0\psi-{\bar\psi}\gamma_1\partial_1\psi
+\frac{1}{2\tau}{\bar\psi}\gamma_0\psi\Big)-M\frac{{\bar\psi}\psi}{(H\tau)^3}\, 
\end{equation}
with $M$ related to the strength of the sine-Gordon interaction.  In passing we may note that an ordinary massive spin-$\frac{1}{2}$ mass term will, as in (\ref{simp_free_lagr_1/2}), will have the mass term divided by $(H\tau)^2$ rather than $(H\tau)^3$. 
The vacuum to vacuum amplitude is 
\begin{equation}\label{vac-vac-1}
\langle 0, {\rm out}|0, {\rm in}\rangle=\exp {\rm tr}  \ln\left(i\slashed{\partial}-M/(H\tau)^3\right)\, ;
\end{equation}
to this end we need the eigenvalues of the Dirac operator, with a nonconstant mass term $\slashed{\partial}-M/(H\tau)^3$. If $\psi$ is an eigenfunction of this operator then $\gamma_5\psi$ is an eigenfunction of $-\slashed{\partial}-M/(H\tau)^3$ with the same eigenvalue and we may replace (\ref{vac-vac-1}) with
\begin{equation}\label{vac-vac-1}
\langle 0, {\rm out}|0, {\rm in}\rangle=\exp {\frac{1}{2}\rm tr}  \ln\left(i\slashed{\partial}-M/(H\tau)^3\right)\left(-i\slashed{\partial}-M/(H\tau)^3\right)\, ,
\end{equation}
which requires us to look at the eigenvalues of $\left(i\slashed{\partial}-M/(H\tau)^3\right)\left(-i\slashed{\partial}-M/(H\tau)^3\right)=
\partial^2+M^2/(H\tau)^6+3i\gamma_0M/(H^3\tau^4)\, .$ After rotating to Euclidian time, $\tau\rightarrow it_{\rm E}$ we want to determine the reality properties of the eigenvalue of the operator (with $e^{ikx}$ spatial dependence and diagonal $\gamma_0$); 
\begin{equation}\label{euclid_oper}
-\partial^2_{t_{\rm E}}+k^2-M^2/(Ht_{\rm E})^6\pm 3iM/(H^3t^4_{\rm E})\, .
\end{equation}
Aside from the explicit imaginary terms, the real part of the above operator is just a one dimensional Schr\"{o}dinger equation with an $1/r^6$ attractive potential resulting in an infinite number  of negative eigenvalues whose logarithms have imaginary parts. The trace in (\ref{vac-vac-1}), after rotating to Euclidian time, introduces an other factor of $i$ resulting in an infinite sum of real contributions to the exponent in (\ref{vac-vac-1}) and a vanishing $\langle 0, {\rm out}|0, {\rm in}\rangle$ amplitude. 

It is instructive to study the ordinary massive spin-$\frac{1}{2}$ action in a de Sitter background. An noted below eq. (\ref{taumass}), terms with $(H\tau)^3$ in the previous discussion vary as $(H\tau)^2$ for the case of an ordinary de Sitter mass term. Following the previous steps results in studying the eigenvalues of 
\begin{equation}\label{euclid_oper-norm}
-\partial^2_{t_{\rm E}}+k^2+M^2/(Ht_{\rm E})^4\pm 2M/(H^2t^3_{\rm E})\, .
\end{equation}
Now the small $t_{\rm E}$ potential is strongly repulsive.  For $t_{\rm E}>M/(2H^2)a$ the potential develops a {\em shallow} attractive part resulting in a {\em finite} number of negative eigenvalues and a 
vacuum to vacuum amplitude that is finite but less than one.  This is a reflection of the thermal particle creation for any field in a curved space-time background.  
\section{Summary}\label{summ}
Even if interacting field theories destabilize an underlying de Sitter space, it is not unreasonable that such interactions involving only fermions do not break de Sitter symmetry or cause the vanishing of $\langle 0|S|0\rangle$; in (1+1) dimensions fermi fields have fewer infrared pathologies than bose ones.  For the cases studied in this work we found that the completely soluble spin-$\frac{1}{2}$ field theories, the Thirring model and massless QED, can be solved by translating them to free boson models with standard mass terms. These do not lead to any instabilities.

On the other hand, starting with an interacting bose field, we found that the equivalent fermion field theory is still free, and thus soluble, but with a mass term that breaks de Sitter symmetry explicitly in that it behaves as $1/\tau$, with $\tau$ being the cosmic time. In addition, the logarithm of the functional determinant governing the vacuum to vacuum amplitude has an infinite number of eigenvalues with imaginary parts resulting in zeros for this amplitude analogous to the result obtained in \cite{ Polyakov:2009nq}.   
\section*{Acknowledgments}
The author wishes to thank Dr. Arvind Rajaraman for many instructive discussions. 



\end{document}